\documentclass[12pt]{article}

\usepackage{amsfonts, amssymb, amscd}
\usepackage{amsmath}
\usepackage[subnum]{cases}

\usepackage{cite}

\def\l{\left}
\def\r{\right}
\def\bl{\Biggl}
\def\br{\Biggr}
\def\nn{\nonumber}
\def\ql{\textquotedblleft}

\def\1o2{{1\over2}}
\def\e={\equiv}
\def\={\neq}
\def\>{\geq}
\def\<{\leq}

\def\a{\alpha}
\def\b{\beta}
\def\ab{{\alpha\beta}}
\def\g{\gamma}

\def\d{\delta}

\def\m{\mu}
\def\n{\nu}
\def\mn{{\mu\nu}}

\def\la{\lambda}
\def\La{\Lambda}
\def\p{\phi}
\def\pa{\partial}

\usepackage{graphicx}

\begin{document}

\title{NAT Black Holes}
\author{ Metin G\"{u}rses$^{(a),(b)}$\footnote{gurses@fen.bilkent.edu.tr},
Yaghoub Heydarzade$^{(a)}$\footnote{yheydarzade@bilkent.edu.tr}, and \c{C}etin \c{S}ent\"{u}rk$^{(c)}$\footnote{csenturk@thk.edu.tr} \\
{\small (a) Department of Mathematics, Faculty of Sciences}\\
{\small Bilkent University, 06800 Ankara, Turkey}\\
 {\small (b) Department of Physics, Faculty of Sciences}\\
{\small Bilkent University, 06800 Ankara, Turkey}\\
{\small (c) Department of Aeronautical Engineering}\\
{\small University of Turkish Aeronautical Association, 06790 Ankara, Turkey}
}


\begin{titlepage}
\maketitle
\thispagestyle{empty}

\begin{abstract}
We study some physical properties of  black holes in Null Aether Theory (NAT)--a vector-tensor theory of gravity. We first review the black hole solutions in NAT and then derive the first law of black hole thermodynamics. The temperature of the black holes depends on both the mass and the NAT \textquotedblleft charge" of the black holes. The extreme cases where the temperature vanishes resemble the extreme Reissner-Nordstr\"{o}m black holes. We also discuss the contribution of the NAT charge to the geodesics of massive and massless particles around the NAT black holes.

\end{abstract}

\end{titlepage}



\setcounter{page}{2}

\newpage

\section{Introduction}

Black holes are of fundamental importance today. This is because of the fact that studies of their properties from both theoretical and observational points of view are being expected to shed much light on the nature of the gravity at strong gravity regimes and at very high energy scales where the gravitational force becomes dominant over the other interactions. For this reason, they have always been at the heart of the theoretical investigations involving gravitational phenomena, especially since the discovery of the four laws of black hole mechanics \cite{bch} and Hawking radiation \cite{haw} in the context of general theory of relativity (GR). More importantly, the thermodynamic interpretation of the four laws \cite{bek} and the attributions of temperature and entropy to black hole horizon have provided  useful information about the nature of quantum gravity through holography \cite{thooft} and its specific realization AdS/CFT correspondence \cite{malda}. Observationally, recent GW events \cite{gws} and the image taken by Event Horizon Telescope Collaboration \cite{eht} have proven the existence of black holes by direct observations, which has also justified the theoretical studies conducted so far.

The event horizon of a black hole is a globally-defined causal boundary which separates the inside of the black hole from the outside. More formally, it is a null surface separating those light rays reaching infinity from those falling to the singularity inside. Since it is defined globally, the determination of the location of the event horizon requires in general the knowledge of the global structure of the spacetime. However, in the case of static spherically symmetric spacetimes, one can introduce convenient coordinate systems in which the determination is made by looking for places where the local light cones tilt over. This implies that the existence of event horizons (and of black holes) has to do with the local Lorentz invariance of the spacetime. Therefore, it is of great importance to explore the properties of black holes in gravity theories that exhibit violations of local Lorentz invariance.


Lorentz symmetry is built in GR which describes gravitation well at low energy scales by assuming the spacetime structure as continuous and smooth, excluding singularities. But this symmetry might be broken at very high energy scales, especially at the Planck or quantum gravity scales, where quantum gravitational effects must be taken into account. In fact, there are theories, such as string theory and loop quantum gravity, contemplating that the quantum fluctuations at or beyond the Planck scale might be so violent that the spacetime ceases to be continuous and has a discrete structure, and thereby the Lorentz symmetry is not valid \cite{matt}. This way of reasoning immediately leads to the contemplation of gravity theories in which Lorentz symmetry is broken explicitly.



One way to construct a Lorentz-violating gravity theory is to assume the existence of a vector field of constant norm which dynamically couples to the metric tensor at each point of spacetime. In other words, the spacetime curvature is determined together by the metric tensor and the coupled vector field in spacetime. Such a vector field is referred to as the \ql aether" because that generally defines a preferred direction in spacetime and breaks the local Lorentz invariance. Eintein-aether theory \cite{jm} is such a theory in which the vector field is timelike everywhere and explicitly breaks the boost sector of the Lorentz symmetry. The internal structure and dynamics of this theory have been studied extensively in the literature \cite{ej1,ej2,gej,tm,bjs,bbm,gs,bs,dww,bsv,gur,gsen,cl,fj,zfs1,bdfsz,zfs2,zzbfs,bl,ab,cdgt,dj,jm1,obw,ghlp}.

Recently, a new vector-tensor theory of gravity called the Null Aether Theory (NAT) \cite{gsen1} has been introduced into the realm of modified gravities. This theory assumes the dynamical vector field (the aether) inherent in the theory to be null at evey point in spacetime. In the paper \cite{gsen1}, we first studied the Newtonian approximation of the theory and showed that it reproduces the Poisson equation at the perturbation order  by, in some cases, rescaling the Newton's constant $G_N$. Then we obtained exact spherically symmetric solutions in this theory by properly choosing the null vector field and we showed that there is a large class of solutions depending on the parameters of the theory. Among these, there are Vaidya-type nonstationary solutions because of the null aether behaving as a null matter source, and for some special values of the parameters, stationary Schwarzschild-(A)dS and Reissner-Nordstr\"{o}m-(A)dS type solutions with some effective cosmological constant and some \ql charge" sourced by the aether, respectively. We also discussed the existence of stationary black holes among these exact solutions for arbitrary values of the parameters of the theory. (See \cite{gsen1} for details and explicit structures of these solutions.) To see the effect of the null aether in cosmology, we studied the flat FLRW metric and, taking the spatial component of null aether lying along the $x$ axis, we found all possible perfect fluid solutions of NAT. We also discussed the existence of the Big-bang singularity and the accelerated expansion of the universe in NAT. In addition to these, to better understand the internal dynamics of the theory, we constructed exact wave solutions by specifically considering the Kerr-Schild-Kundt (KSK) class of metrics \cite{ggst,ghst} with maximally symmetric backgrounds. After giving the exact AdS-plane wave solutions of NAT in $D\>3$ dimensions, we also obtained all possible $pp$-wave solutions of the theory propagating in the flat background spacetime. These exact wave solutions are consistent with the linearized waves of the theory \cite{gsen2}.

In this paper, we will continue our explorations in the implications of the exact spherically symmetric solutions and black hole spacetimes found in \cite{gsen1}. After giving a brief review of the Newtonian limit and static spherically symmetric solutions of NAT, we will first discuss the possible effect of the null aether field on the solar system dynamics by extracting the so-called Eddington-Robertson-Schiff parameters $\b$ and $\g$ for our solutions, which explicitly appear in the perihelion precession and the light deflection expressions. We will see that, at the post-Newtonian order, there is no contribution from the aether field to the deflection of light rays passing near a massive body; that is the same as in GR! However, there is an explicit contribution, at the post-Newtonian order, from the aether field to the perihelion precession of planetary orbits. This fact can be used to constrain the parameters of the theory from solar system observations. Then we shall present the details of the black hole spacetimes by discussing the singularity structure, the ADM mass of the asymptotically flat solutions, and the thermodynamics in order. In the thermodynamics of NAT black holes especially, it is interesting to note that an appropriate definition of the NAT \ql charge" reduces the horizon thermodynamics to that of the Reissner-Nordstr\"{o}m-(A)dS black hole in GR and the first law takes the standard form if the theory's parameters $c_2$ and $c_3$ satisfy a strict condition. Lastly, we will also discuss the circular geodesics of massive and massless particles around the NAT black holes to see the effect of the null aether on the particle trajectories in the spacetime. We will show that the null aether substantially changes the behavior of the circular orbits of massive and massless particles. We will also calculate the perihelion precession of planets and the deflection of light rays explicitly in the case of a nonzero cosmological constant.

The organization of the paper is as follows. In Sec. 2, we give the Null Aether Theory in detail. In Sec. 3 we review the Newtonian approximation of the theory and observe that the results we obtained in this section are consistent with the exact solutions in the next section. In Sections 4 and 5, we discuss exact spherically symmetric solutions and black hole spacetimes in NAT, respectively. In Sec. 6, we obtain the ADM mass of the asymptotically flat NAT black holes. In Sec. 7, we study the first law of black hole thermodynamics. In Sec. 8, we obtain the circular orbits of massive and massless particles around the NAT black holes.

We use the metric signature $(-,+,+,+,\ldots)$ throughout the paper.

\section{Null Aether Theory}

Aether theory is a generally covariant theory of gravity in which the metric tensor ($g_\mn$) of the spacetime dynamically couples, through covariant derivatives, to a vector field ($v^\m$)--referred to as the \ql aether." In the absence of matter fields, the action of the theory can be written as \cite{gsen1}
\begin{equation}
I={1 \over 16 \pi G}\, \int d^{4}x\sqrt{-g}\,[R-2\Lambda-K^{\m\n}\,_{\a\b}\nabla_{\m}v^{\a}\nabla_{\n}v^{\b}
+\lambda(v_{\mu}v^{\mu}+\varepsilon)],\label{action}
\end{equation}
where $R$ is the Ricci scalar, $\La$ is the bare cosmological constant, and
\begin{equation}
K^{\mu \nu}\,_{\alpha \beta}=c_{1}g^{\mu \nu}g_{\alpha
\beta}+c_{2}\delta^{\mu}_{\alpha}
\delta^{\nu}_{\beta}+c_{3}\delta^{\mu}_{\beta}
\delta^{\nu}_{\alpha}-c_{4}v^{\mu}v^{\nu}g_{\alpha
\beta},\label{Ktensor}
\end{equation}
with the dimensionless constant parameters $c_{1}$, $c_{2}$, $c_{3}$, $c_{4}$. From now on, throughout the text, we shall use the shorthand notation $c_{ij}=c_i+c_j$ for combinations of these constants. When $\varepsilon=-1$, the aether field is timelike and this case corresponds to the Einstein-Aether theory of \cite{jm}. In our case, however, $\varepsilon=0$ and the aether becomes a null vector field. The Lagrange multiplier $\la$ in (\ref{action}) is introduced into the theory to explicitly enforce the nullity of the vector field; that is, to have
\begin{equation}\label{con}
v_{\mu}v^{\mu}=0
\end{equation}
at each point of the spacetime. Therefore the independent variables in the theory are $g^\mn$, $v^\m$, and $\la$. The field equations are then obtained by varying the action (\ref{action}) with respect to these fields: Varying with respect to $\lambda$ immediately leads to the null constraint (\ref{con}) and, making use of it, varying with respect to $g^{\mu\nu}$ and $v^\mu$ respectively yields
\begin{eqnarray}
&&G_{\mu \nu}+\Lambda g_{\mu\nu}=\nabla_{\alpha}\l[ J^{\alpha}\,_{(\mu}
\,v_{\nu)}-J_{(\mu}\,^{\alpha}\, v_{\nu)}+J_{(\mu \nu )}\,
v^{\alpha}\r] \nonumber\\
&&~~~~~~~~~~~~~~~~~~+c_{1}\l(\nabla_{\mu}v_{\alpha}\nabla_{\nu}
v^{\alpha}-\nabla_{\alpha}v_{\mu}\nabla^{\alpha}
v_{\nu}\r) \nonumber \\
&&~~~~~~~~~~~~~~~~~~+c_{4}\dot{v}_{\mu}\dot{v}_{\nu}+\lambda v_{\mu}v_{\nu}-{1 \over 2} L g_{\mu \nu}, \label{eqn01}\\
&&\nonumber\\
&&c_{4} \dot{v}^{\alpha} \nabla_{\mu}
v_{\alpha}+\nabla_{\alpha}J^{\alpha}\,_{\mu}+\lambda
v_{\mu}=0,\label{eqn02}
\end{eqnarray}
where we used the identifications
\begin{eqnarray}
&&\dot{v}^{\mu}\equiv v^{\alpha}\nabla_{\alpha}v^{\mu},\\
&&J^{\mu}\,_{\a}\equiv K^{\mu \n}\,_{\a \b}\nabla_{\n}v^{\beta},\label{J}\\
&&L\equiv J^{\mu}\,_{\a}\nabla_{\mu}v^{\a}.\label{L}
\end{eqnarray}
Obviously, the Minkowski metric ($\eta_\mn$) together with a constant null vector ($v_\m=const.$) and $\la=0$ constitute a solution to NAT.
Since being null, the zero ather field (i.e. $v_\m=0$) with an arbitrary $\la$ reduces the theory to the usual general relativity; however, this trivial case can be distinguished from the nontrivial aether case by imposing certain initial and boundary conditions on the solutions of the Einstein-Aether equations (\ref{eqn01}) and (\ref{eqn02}). (See the discussion in \cite{gsen1}.)

Since the aether field in NAT is null by construction, one can always introduce a scalar degree of freedom into the theory. The reasoning is as follows: First set up, at each point in spacetime, a null tetrad $e^a_\m=(l_\m,n_\m,m_\m,\bar{m}_\m)$, where $l_\m$ and $n_\m$ are real null vectors with $l_\m n^\m=-1$, and $m_\m$ is a complex null vector orthogonal to $l_\m$ and $n_\m$, and then assume the null aether $v_\m$ is proportional to the one null leg of this tetrad, say $l_\m$; i.e. $v_{\mu}=\phi(x)l_{\mu}$. Thus this geometric construction enables us to naturally introduce a scalar function $\p(x)$--the spin-0 part of the aether field--which generally contains the physical meaning of the aether by carrying a nonzero \ql aether charge."

\section{Newtonian Limit of Null Aether Theory}

The Newtonian limit of the theory can be achieved by assuming the gravitational field is weak and static and produced by a nonrelativistic matter field. Also the cosmological constant plays no role in this context so that it can be set equal to zero. Therefore in taking the Newtonian limit, we can write the metric in $x^\m=(t,x,y,z)$ as
\begin{equation}\label{newt}
  ds^2=-[1+2\Phi(\vec{x})]dt^2+[1-2\Phi(\vec{x})](dx^2+dy^2+dz^2),
\end{equation}
where $\Phi(\vec{x})$ is the gravitational potential on the order of $G$, and take the matter energy-momentum tensor as
\begin{equation}\label{EMgen}
T^{matter}_{\mu \nu}=(\rho_m+p_m) u_{\mu} u_{\nu}+p_m g_{\mu \nu}+t_{\mu \nu},
\end{equation}
where $u_{\mu}=\sqrt{1+2\Phi}\, \delta_{\mu}^{0}$ is the four-velocity of the matter field, $\rho_m$ and $p_m$ are the mass density and pressure, and $t_{\mu \nu}$ is the stress tensor
with $u^{\mu} t_{\mu \nu}=0$. Then perturbing also the aether field appropriately, we consider only the zeroth and first order (linear) terms in $v_\m$ and $g_\mn$ in the Eistein-Aether equations (\ref{eqn01}) and (\ref{eqn02}). At this point, however, there appear three distinct cases in perturbing the aether field, with the associated Newtonian limits:

\vspace{0.3cm}
\noindent
{\bf Case 1:} Let us decompose the null aether field as
\begin{equation}
v_{\mu}=a_{\mu}+k_{\mu},
\end{equation}
where $a_{\mu}=(a_{0},a_{1},a_{2},a_{3})$ is a constant null vector representing the background aether field and $k_{\mu}=(k_{0},k_{1},k_{2},k_{3})$ is the perturbation which need not necessarily be a null vector. The null constraint (\ref{con}) then implies that
\begin{eqnarray}
&&a_{0}^2=\vec{a} \cdot \vec{a}, \\
&&k_{0}=\frac{1}{a_{0}}\,[\vec{a} \cdot \vec{k}+2a_{0}^2\Phi],
\end{eqnarray}
at the perturbation order. Since the metric is symmetric under rotations, we can take, without loosing any generality, $a_{1}=a_{2}=0$
and for simplicity we will assume that $k_{1}=k_{2}=0$. Then one can show that
\begin{eqnarray}
&&c_{3}=-c_{1}=-c_{2},~~k_{3}=-\frac{2 a_{3}^3 c_{4}}{c_{1}}\,\Phi,\\
&&\nn\\
&&\nabla^2\, \Phi=\frac{4 \pi G}{1-c_{1}\,a_{3}^2}\, \rho_{m}=4 \pi G_{N} \rho_{m}.\label{case12}
\end{eqnarray}
The last equation (\ref{case12}) is in the form of the Poisson equation and implies that Newton's gravitation constant $G_N$ is an effective one defined by the scaling
\begin{equation}
G_{N}=\frac{G}{1-c_{1}\,a_{3}^2}.
\end{equation}
Similar scaling also appears in the context of Einstein-Aether theory \cite{cl,fj}. The constraint $c_{3}+c_{1}=0$ can be removed by taking the stress part $t_{\mu \nu}$ into account in the energy momentum tensor, then there remains only the constraint $c_{2}=c_{1}$.

\vspace{0.3cm}
\noindent
{\bf Case 2:} Take the null aether field as $v_\m=\p(\vec{x})l_\m$ where $l_\m$ is a null vector defined by the geometry (\ref{newt}) as
\begin{equation}\label{}
  l_\m=\d_\m^0+(1-2\Phi)\frac{x^i}{r}\d_\m^i,
\end{equation}
with $r=\sqrt{x^2+y^2+z^2}$ and $i=1,2,3$. Note that any multiplicative function of $\vec{x}$ can be absorbed into the scalar function $\p(\vec{x})$. Now assuming the perturbation $\p(\vec{x})=\phi_{0}+\phi_{1}(\vec{x})$ where $\phi_{0}=const.\neq0$ and $\phi_{1}$ is at the same order as $G$, we obtain
\begin{eqnarray}
&&c_{1}+c_{3}=0,~~c_{2}=0,~~c_{4}=0,~~\phi_{1}=2\phi_0 \Phi,\\
&&\nn\\
&&\nabla^2 \Phi=\frac{4 \pi G}{1-c_{1}\phi_{0}^2}\,\rho_m=4 \pi G_{N}\rho_m.\label{case22}
\end{eqnarray}
Again, the effective value of Newton's constant can be seen from (\ref{case22})
\begin{equation}\label{}
  G_{N}=\frac{G}{1-c_{1}\phi_{0}^2}.
\end{equation}
This is, however, a very restricted aether theory because there is only one independent parameter $c_{1}$ left in the theory.

\vspace{0.3cm}
\noindent
{\bf Case 3:} Take the zeroth order scalar aether field in Case 2 as zero; i.e., $\phi_{0}=0$. This means that $\p(\vec{x})=\phi_{1}(\vec{x})$ and is at the same order as $G$. Therefore, there is no contribution to the equation (\ref{eqn01}) from the aether field at the linear order in $G$, and from the 00 component of (\ref{eqn01}), we get
\begin{equation}\label{Pois}
  \nabla^2\Phi=4\pi G\rho_m,
\end{equation}
which is the Poisson equation unaffected by the null aether field at the perturbation order. On the other hand, from the $i$th component of the aether equation (\ref{eqn02}) we obtain, at the linear order in $G$,
\begin{eqnarray}
&&(c_2+c_3)r^2x^j\pa_j\pa_i\p-(2c_1+c_2+c_3)x^ix^j\pa_j\p\nn\\
&&~~~~~~~~+[2c_1+3(c_2+c_3)]r^2\pa_i\p-2(c_1+c_2+c_3)x^i\p=0,\label{Newcon}
\end{eqnarray}
after eliminating the Lagrange multiplier $\la$ by using the zeroth order equation.


In the case of spherical symmetry, outside the mass distribution of mass $M$, the Poisson equation (\ref{Pois}) gives
\begin{equation}\label{}
  \Phi(r)=-\frac{GM}{r},
\end{equation}
and the condition (\ref{Newcon}) gives
\begin{equation}\label{philin}
 \p(r)=\frac{a_1}{r^{(1+q)/2}}+\frac{a_2}{r^{(1-q)/2}},
\end{equation}
where $a_1$ and $a_2$ are arbitrary constants on the order of $G$ and we have defined the parameter
\begin{equation}\label{}
q\e=\sqrt{9+8\frac{c_1}{c_{23}}},
\end{equation}
which is always positive by definition. Therefore, we can immediately see that the three of the parameters of NAT must satisfy the constraint
\begin{equation}\label{}
  \frac{c_1}{c_{23}}\geq-\frac{9}{8}.
\end{equation}
Specifically, when $q=0$ ($c_1=-9c_{23}/8$), we have
\begin{equation}\label{}
  \p(r)=\frac{a_1+a_2}{\sqrt{r}};
\end{equation}
when $q=3$ ($c_1=0$), we have
\begin{equation}\label{}
  \p(r)=\frac{a_1}{r^{2}}+a_2r;
\end{equation}
or when $q=1$ ($c_1=-c_{23}$), we have
\begin{equation}\label{q=1}
  \p(r)=\frac{a_1}{r}+a_2.
\end{equation}

\section{Spherically Symmetric Static Solutions in Null Aether Theory}

In this section, we shall review the spherically symmetric static solutions in NAT found previously in the original work \cite{gsen1}. The metric written in the Eddington-Finkelstein coordinates $x^{\mu}=(u,r,\theta, \varphi)$ is
\begin{equation}\label{BHKS}
  ds^2=-\l[1-\frac{\La}{3}r^2-2f(r)\r]du^2+2dudr+r^2d\theta^2+r^2\sin^2\theta d\varphi^2,
\end{equation}
where $u$ is the null coordinate, then taking the null aether field--assumed to be present at each spacetime point in the theory--is aligned with this coordinate, we obtain the solution
\begin{eqnarray}
& v_\m=\p(r)\delta^{u}_\m,\\
&\nn\\
& \displaystyle \p(r)=\frac{a_1}{r^{(1+q)/2}}+\frac{a_2}{r^{(1-q)/2}}, \label{phiexact}\\
&\nn\\
& f(r)=\left\{\begin{array}{ll}
         \displaystyle \frac{a_1^2b_1}{r^{1+q}}+\frac{a_2^2b_2}{r^{1-q}}+\frac{\tilde{m}}{r},
         &\mbox{for $\displaystyle q\neq0$,}\\
         &\\
         \displaystyle \frac{m}{r},
         &\mbox{for $\displaystyle q=0$,}\label{fm}
\end{array} \right.
\end{eqnarray}
where $a_1$, $a_2$, $\tilde{m}$, and $m$ are just integration constants and
\begin{equation}\label{qbb}
  q\equiv\sqrt{9+8\frac{c_1}{c_{23}}},~~b_1=\frac{1}{8}[c_3-3c_2+c_{23}q],~~b_2=\frac{1}{8}[c_3-3c_2-c_{23}q].
\end{equation}
As we will show later, the constants $\tilde{m}$ and $m$ are the mass parameters of the solutions. At this point, it is also important to note that the exact solution (\ref{phiexact}) is the same as the linearized one (\ref{philin}) obtained in the previous section. This means that the null aether contribution to the metric [see Eq. (\ref{fm})] comes in at the order of $G^2$.

Now performing the coordinate transformation
\begin{equation}\label{utTrans}
du=dt+\frac{dr}{1-\frac{\La}{3}r^2-2f(r)},
\end{equation}
one can bring the metric (\ref{BHKS}) into the Schwarzschild coordinates
\begin{equation}\label{BHSch}
  ds^2=-h(r)dt^2+\frac{dr^2}{h(r)}+r^2d\theta^2+r^2\sin^2\theta d\varphi^2,
\end{equation}
where
\begin{equation}\label{}
  h(r)\equiv1-\frac{\La}{3}r^2-2f
  =\left\{\begin{array}{ll}
         \displaystyle 1-\frac{\La}{3}r^2-\frac{2a_1^2b_1}{r^{1+q}}-\frac{2a_2^2b_2}{r^{1-q}}-\frac{2\tilde{m}}{r}
         &\mbox{(for $q\neq0$),}\\
         &\\
         \displaystyle 1-\frac{\La}{3}r^2-\frac{2m}{r}
         &\mbox{(for $q=0$).}\label{hm}
\end{array} \right.
\end{equation}
This metric describes the spherically symmetric static solutions in NAT, and interestingly we have lots of them due to the free parameters $q$, $b_1$, and $b_2$ in the theory. The solution for $q=0$ is the usual Schwarzschild-(A)dS spacetime but there are also solutions corresponding to some other specific values of the parameter $q$ which are of special importance; for instance,
\begin{itemize}
  \item When $q=1$ ($c_1=-c_{23}$), $h(r)\equiv1-A-\La r^2/3-B/r^2-2\tilde{m}/r$, where $A\e=2a_2^2b_2$ and $B\e=2a_1^2b_1$: This is a Reissner-Nordstr\"{o}m-(A)dS type solution if $A=0$.
  \item When $q=2$ ($c_1=-5c_{23}/8$), $h(r)\equiv1-\La r^2/3-A/r^3-Br-2\tilde{m}/r$, where $A\e=2a_1^2b_1$ and $B\e=2a_2^2b_2$: This solution with $A=0$ has been obtained by Mannheim and Kazanas \cite{mk} in conformal gravity who also argue that the linear term $Br$ can explain the flatness of the galaxy rotation curves.
  \item When $q=3$ ($c_1=0$), $h(r)\equiv1-A/r^4-Br^2-2\tilde{m}/r$, where $A\e=2a_1^2b_1$ and $B\e=\La/3+2a_2^2b_2$: This is a Schwarzschild-(A)dS type solution if $A=0$. Solutions involving terms like $A/r^4$ can be found in, e.g., \cite{bbm,gser}.
\end{itemize}

Before concluding this section, one last remark must be made on the possible effects of the null aether field on the solar system observations. For this purpose, we will consider the post-Newtonian parameters in the case of a static, spherically symmetric mass distribution like the Sun. Since the cosmological constant is totally negligible in this setting, the metric produced by such a body can be expanded to post-Newtonian order as \cite{wein}
\begin{eqnarray}
&&ds^2=-\l(1-\frac{2GM}{r}+2(\b-\g)\frac{G^2M^2}{r^2}+\cdots\r)dt^2\nn\\
&&~~~~~~~~+\l(1+2\g\frac{GM}{r}+\cdots\r)dr^2+r^2d\theta^2+r^2\sin^2\theta d\varphi^2,\label{PPN}
\end{eqnarray}
where $M$ is the mass of the body and $\b$ and $\g$ are the so-called Eddington-Robertson-Schiff parameters. These two parameters explicitly appear in the expressions for the perihelion precession of a planetary orbit and the deflection of light rays passing near the body which are respectively given by
\begin{eqnarray}
&&\d\varphi=\l(\frac{2-\b+2\g}{3}\r)\frac{6\pi GM}{a(1-e^2)},\label{peri}\\
&&\nn\\
&&\d\psi=\l(\frac{1+\g}{2}\r)\frac{4GM}{b},\label{light}
\end{eqnarray}
where $a$ is the semi-major axis and $e$ is the eccentricity of the orbit and $b$ is the impact parameter. In general relativity, from the Schwarzschild metric, it can immediately be seen that $\b=\g=1$.

In NAT, we have the solutions given by (\ref{BHSch}) and (\ref{hm}). So taking $\La=0$, for the case $q=0$, since we recover the usual Schwarzschild solution, we can immediately have $\b=\g=1$ just as in GR, but when $q$ is a positive integer, the expanded metric is
\begin{eqnarray}\label{}
&&ds^2=-\l(1-\frac{2\tilde{m}}{r}-\frac{2a_1^2b_1}{r^{2}}+\cdots\r)dt^2\nn\\
&&~~~~~~~~+\l(1+\frac{2\tilde{m}}{r}+\cdots\r)dr^2+r^2d\theta^2+r^2\sin^2\theta d\varphi^2,\label{NATPPN}
\end{eqnarray}
where we have assumed $a_2=0$ just for simplicity. It should be noted that the terms with $q>1$ do not contribute to the post-Newtonian order. In other words, only the term with $q=1$ has contribution to the post-Newtonian order. Now, knowing that $\tilde{m}\sim G$ and $a_1\sim G$ and comparing (\ref{NATPPN}) with (\ref{PPN}), we can read off the post-Newtonian parameters as
\begin{equation}\label{}
  \b=1-\frac{a_1^2b_1}{\tilde{m}^2},~~\g=1.
\end{equation}
Therefore, we can see from (\ref{light}) that the null aether does not affect the light deflection at the post-Newtonian order; it is the same as in GR. However, it is obvious from (\ref{peri}) that it does affect the perihelion precessions of planets as
\begin{equation}\label{deltaphi1}
  \d\varphi=\l(1+\frac{a_1^2b_1}{3\tilde{m}^2}\r)\frac{6\pi \tilde{m}}{a(1-e^2)},
\end{equation}
This result tells us that, if $b_1>0$, the perihelion advance is greater than that of GR, and if $b_1<0$, it is less than that of GR.

\section{Black Hole Solutions in Null Aether Theory}

The metric (\ref{BHSch}) also describes spherically symmetric static black hole solutions in NAT. The event horizons of these solutions are in principle determined by the positive real roots of the equation $h(r)=0$ [see Eq. (\ref{hm})]. In general, the existence of these roots crucially depends on the signs and/or values of and the relation between the parameters $(q, \La, a_1, a_2, b_1, b_2, \tilde{m}, m)$ appearing in (\ref{hm}). For example, in the case $q=0$, there are two distinct positive real roots, which are those of the usual Schwarzschild-dS black hole, if $\La>0$ and $0<9\La m^2<1$, and there is only one positive root, which is that of the usual Schwarzschild-AdS black hole, if $\La<0$. On the other hand, the determination of the positive real roots of the equation $h(r)=0$ in the other case $q\=0$ is not that easy. However, we can generally make the following points. If $q$ is an integer, $h(r)=0$ becomes a polynomial equation which may have at least one positive real root representing the event horizon of the corresponding black hole. And, if $q$ is not an integer, the limits $\lim_{r\rightarrow0^+}h(r)$ and $\lim_{r\rightarrow\infty}h(r)$ may be used to just determine the existence of the real roots; more explicitly, since $h(r)$ is a continuous function of $r$, when the signs of the limits are opposite, it is certain that there is at least one real root of $h(r)$. For example, in Table (1), we classified
the cases in which there is at least one real root of the equation $h(r)=0$. There might be other possibilities, of course, but by giving these examples, we are trying to point out that there are black hole solutions in the general case $q\=0$ as well.
\begin{table}[!ht]
\centering
\begin{tabular}{|c|c|c|c|c|c|} 
\hline\hline
$q$ & $b_1$& $b_2$ & $\Lambda$& $\lim_{r\rightarrow0^+}h(r)$& $\lim_{r\rightarrow\infty}h(r)$
\\ [0.5ex]
\hline 
$(0,3)$&+&$\pm$ &-&-&+\\[2ex]
$(0,3)$&-&$\pm$ &+&+&-\\[2ex]
$(3, \infty)$  &  $+$  & $-$ & $\pm$  & $-$ & $+$\\[2ex]
$(3, \infty)$  &  $-$  & $+$ & $\pm$  & $+$ & $-$\\[2ex]
\hline\hline 
\end{tabular}
\label{tab:table}
\caption{Some cases in which black holes certainly exist in NAT.}
\end{table}

Black hole solutions may have one or multiple horizons. We call $r=r_{0}$ the largest root of $h(r)$ and hence the one corresponding to the event horizon.
When there is only one event horizon, the metric function $h(r)$ can be written as
\begin{equation}\label{}
  h(r)=(r-r_0)g(r),
\end{equation}
where $g(r)$ is a continuous function for $r\geq r_0$ and $g(r)>0$ because $h(r)$ must be positive for $r>r_0$. This means that
\begin{equation}\label{}
  h'(r_0)=g(r_0)>0
\end{equation}
due to the continuity of $g(r)$. When there are multiple event horizons, say the number is $m$, the metric function $h(r)$ should be in the form
\begin{equation}\label{hmul}
  h(r)=(r-r_1)(r-r_2)\ldots(r-r_m)g(r),
\end{equation}
where $g(r)>0$ for $r$ greater than the largest root, say $r_0$. Again, due to the continuity of $g(r)$ for $r\geq r_0$,
\begin{equation}\label{}
  h'(r_0)=g(r_0)>0,
\end{equation}
where we assume that all the roots are distinct and the event horizon is at $r_0$, the largest root of (\ref{hmul}).
When some or all of the roots are coincident, we have the extreme case. For example, for two coincident roots,
\begin{equation}\label{}
  h(r)=(r-r_0)^2g(r),
\end{equation}
where $g(r)>0$ for $r>r_0$. Then
\begin{equation}\label{}
  h'(r_0)=0.
\end{equation}
From now on, we shall admit this condition as the indicator of an extreme black hole.

To understand the singularity structure of our solutions given in (\ref{BHSch}) and (\ref{hm}), we shall calculate the two of the curvature scalars; namely, the Ricci and Kretschmann scalars. For $q\=0$, they are
\begin{eqnarray}
&& R=4\La+2q\l[\frac{A_1(q-1)}{r^{3+q}}+\frac{A_2(q+1)}{r^{3-q}}\r],\\
&&\nn\\
&&K=R_{\mn\ab}R^{\mn\ab}\nn\\
&&~~~=\frac{48\tilde{m}^2}{r^6}+\frac{8\La^2}{3}+\frac{8q\La}{3}\l[\frac{A_1(q-1)}{r^{3+q}}+\frac{A_2(q+1)}{r^{3-q}}\r]\nn\\
&&~~~~~+16\tilde{m}\l[\frac{A_1(q+2)(q+3)}{r^{6+q}}+\frac{A_2(q-2)(q-3)}{r^{6-q}}\r]\nn\\
&&~~~~~+4\bl[\frac{A_1^2(12+20q+17q^2+6q^3+q^4)}{r^{2(3+q)}}+\frac{2A_1A_2(12-9q^2+q^4)}{r^{6}}\nn\\
&&~~~~~~~~~~~~~~~~~~~~~~+\frac{A_2^2(12-20q+17q^2-6q^3+q^4)}{r^{2(3-q)}}\br],
\end{eqnarray}
where we made the definitions $A_1\e=a_1^2b_1$ and $A_2\e=a_2^2b_2$. It can be seen that the only singularity is at $r=0$. From these, we can also recover the standard Schwarzschild-(A)dS expressions by setting $A_1=0$ and $A_2=0$ simultaneously.

\section{ADM Mass of Asymptotically Flat Solutions}

To obtain asymptotically flat solutions, we should immediately take $\La=0$, and the metric (\ref{BHSch}) becomes
\begin{equation}\label{}
  ds^2=-h(r)dt^2+\frac{dr^2}{h(r)}+r^2d\theta^2+r^2\sin^2\theta d\varphi^2,
\end{equation}
where
\begin{equation}\label{}
  h(r)=\left\{\begin{array}{ll}
         \displaystyle 1-\frac{2a_1^2b_1}{r^{1+q}}-\frac{2a_2^2b_2}{r^{1-q}}-\frac{2\tilde{m}}{r}
         &\mbox{(for $q\neq0$),}\\
         &\\
         \displaystyle 1-\frac{2m}{r}
         &\mbox{(for $q=0$).}\label{}
\end{array} \right.
\end{equation}
As is obvious, in the $q=0$ case, the metric is just the usual Schwarzschild spacetime which is explicitly asymptotically flat. However, in the $q\neq0$ case, to achieve asymptotically flat boundary condiaitions, one should consider the following cases separately: Since $q>0$ by definition (see Eq. (\ref{qbb})),
\begin{equation}\label{}
  h(r)\mid_{r\rightarrow\infty}=1 \left\{\begin{array}{ll}
         \displaystyle  \mbox{for $0<q<1$}
         &\mbox{(if $a_1\neq0$ and $a_2\neq0$) or (if $a_1=0$ or $b_1=0$),}\\
         &\\
         \displaystyle  \mbox{for $0<q$}
         &\mbox{(if $a_2=0$ or $b_2=0$).}\label{hinf}
\end{array} \right.
\end{equation}

For stationary spacetimes with the time translation Killing vector $\chi^\mu$, the ADM and Komar masses are identical. So, the ADM mass can be calculated from
\begin{equation}\label{adm}
M_{ADM}=-\frac{1}{4\pi G}\int_{\mathcal{B}_\infty}\nabla^\mu\chi^\nu
d\Sigma_{\mu\nu},
\end{equation}
where $d\Sigma_{\mu\nu}=-u_{[_a} s_{_b]}dA$, with $dA=r^2\sin\theta d\theta d\varphi$, is the differential surface element on a two-sphere $\mathcal{B}$ living in a spacelike hypersurface $\Sigma$ of the spacetime. Here, $u_\mu=-\sqrt{h}\delta^t_\mu$ and $s_\mu=\delta^r_\mu/\sqrt{h}$ are the unit timelike and spacelike normals to $\mathcal{B}$, respectively, and $\mathcal{B}_\infty$ is a two-sphere at spatial infinity. Regarding the stationary nature of our spacetime (\ref{BHSch}), the corresponding Killing vector field is $\chi^\mu=\delta^{\mu}_{t}$ and
\begin{equation}
\nabla^\mu\chi^\nu
d\Sigma_{\mu\nu}=-\frac{h'}{2}dA,
\end{equation}
where $h(r)$ is given by (\ref{hm}) with $\La=0$ and the prime denotes differentiation with respect to $r$. Then, the ADM mass in (\ref{adm}) reduces to
\begin{equation}
M_{ADM}=\frac{r^2}{2G} h' \mid_{r\rightarrow\infty}.
\end{equation}
For the case $q=0$, the ADM mass reads as
\begin{equation}
M_{ADM}=\frac{m}{G},
\end{equation}
but for the case $q\neq0$, we obtain
\begin{equation}
M_{ADM}=\frac{1}{G}\l[ \tilde m +(1+q)\frac{a_1^2 b_1}{r^q}+(1-q)\frac{a_2^2 b_2}{r^{-q}} \r]\mid_{r\rightarrow
\infty}.
\end{equation}
Then one realizes that, for having an asymptotically well defined ADM mass for NAT black holes,
\begin{equation}\label{}
 M_{ADM}=\frac{\tilde m }{G}\left\{\begin{array}{ll}
         \displaystyle  \mbox{for $0<q<1$}
         &\mbox{(if $a_1=0$ or $b_1=0$),}\\
         &\\
         \displaystyle \mbox{for $0<q$}
         &\mbox{(if $a_2=0$ or $b_2=0$).}\label{adm1}
\end{array} \right.
\end{equation}
In all these cases, the ADM mass is rescaled through the definition of $G$ in the theory; for example, in the Newtonian limit Case 1 of the previous section, $G=G_N(1-c_1 a_3^2)$ and
\begin{equation}
M_{ADM}=\frac{\tilde m }{G_N(1-c_1 a_3^2)}.
\end{equation}

Although  both cases $a_2=0$ and $b_2=0$ give the same ADM mass (\ref{adm1}) for $q>0$ for
an observer at infinity, they differ if one considers the aether field $\phi$ by putting different constraints on the parameter $q$. That is,
\\
\begin{equation}
\left\{
\begin{array}{lr}
\mbox{If $a_2 =0$}~~\Rightarrow~~\displaystyle\phi=\frac{a_1}{r^{(1+q)/2}},~~0<q, \\\\
\mbox{If $b_2 =0$}~~\Rightarrow~~ \displaystyle \phi=\frac{a_1}{r^{(1+q)/2}}+\frac{a_2}{r^{(1-q)/2}},~~0<q=\frac{c_3 - 3c_2}{c_{23}}<1.
\end{array}
  \right.
\end{equation}
\\
For both of these cases, the constraints on $q$ parameter  guarantees that the aether field is also well behaved at asymptotic region.

\section{Thermodynamics of NAT Black holes}

Now we shall study the thermodynamics of NAT black holes that we reviewed in Sec. 5. Here we first consider the case $a_2=0$. Then the metric function $h(r)$ and the scalar aether field $\p(r)$ take the forms
\begin{eqnarray}
&&h(r)=1-\frac{\La}{3}r^2-\frac{2a_1^2b_1} {r^{1+q}}-\frac{2\tilde{m}}{r},\label{hLa}\\
&&\p(r)=\frac{a_1}{r^{(1+q)/2}}.\label{phiLa}
\end{eqnarray}
The location of the event horizon $r_0$ is given by $h(r_0)=0$ and the area of the event horizon is $A=4\pi r_0^2$. Now let $a_1=GQr_0^{(q-1)/2}$, where $Q$ is the NAT \ql charge." With this identification, (\ref{hLa}) and (\ref{phiLa}) become
\begin{eqnarray}
&&h(r)=1-\frac{\La}{3}r^2-\frac{2G^2Q^2b_1} {r^2}\l(\frac{r_0}{r}\r)^{q-1}-\frac{2\tilde{m}}{r},\label{hLa1}\\
&&\p(r)=\frac{GQ}{r}\l(\frac{r_0}{r}\r)^{(q-1)/2}.\label{}
\end{eqnarray}
At the event horizon location $r_0$, we then have
\begin{eqnarray}
&&h(r_0)=1-\frac{\La}{3}r_0^2-\frac{2G^2Q^2b_1} {r_0^2}-\frac{2\tilde{m}}{r_0}=0,\label{hLa2}\\
&&\p(r_0)=\frac{GQ}{r_0}.\label{phiLa2}
\end{eqnarray}
It is interesting that the horizon condition (\ref{hLa2}) is independent of the parameter $q$ and, when $b_1\e=\frac{1}{8}[c_3-3c_2+c_{23}q]=-1/2$, it becomes that of the Reissner-Nordstrom-(A)dS black hole in GR. In addition, the scalar aether field $\p(r)$ resembles the electric potential at $r=r_0$.

Now assuming the entropy as $S=kA$, where $k$ is a positive constant which takes the value $1/4$ \cite{haw}, and varying that, we obtain
\begin{equation}\label{deltaS}
\delta S=8\pi kr_0\left( r_{0\tilde m}\delta \tilde m + r_{0Q} \delta Q+r_{0\La} \delta \La\right),
\end{equation}
where $r_{0\tilde m}=\frac{\partial r_0}{\partial m}$, $r_{0Q}=\frac{\partial r_0}{\partial Q}$, and $r_{0\La}=\frac{\partial r_0}{\partial \La}$. This relation can be translated into the form of  the first law of thermodynamics as
\begin{equation}\label{1stLaw}
\frac{\delta \tilde m}{G}=T\delta S+V_{\phi}\delta Q+V\delta P,
\end{equation}
where the temperature $T$, the NAT charge potential $V_\phi$, the event horizon volume $V$, and the pressure $P$ are given by
\begin{eqnarray}
T&=&\frac{1}{8\pi G k r_0 r_{0\tilde m}}=\frac{1}{16\pi G k}h^\prime (r_0),\label{temp} \\
V_\phi&=&-\frac{1}{G}\frac{r_{0Q}}{r_{0\tilde m}}=-2b_1\frac{GQ}{r_0}=-2b_1\phi (r_0)\label{V},\\
V&=&8\pi\frac{r_{0\La}}{r_{0\tilde
m}}=\frac{4}{3}\pi r_0^3\label{Vol},\\
P&=&-\frac{\La}{8\pi G},\label{P}
\end{eqnarray}
where $b_1$ takes $-1/2$ to get the standard expression for the fist law. By using the discussions in Sec. 5, we can now explicitly see from (\ref{temp}) that $T>0$ for the non-extreme cases and $T=0$ for all the extreme cases.

As a remark, it is worth mentioning the following point. The extremal event horizon $r_0$ is a radius where $h(r_0)=0$ and $h^\prime(r_0)=0$, and so, when $\La=0$, the extremal event horizon for (\ref{hLa2}) can be obtained as
\begin{equation}
r_0=\tilde m,
\end{equation}
which can equivalently be written in terms of mass and aether charge as
\begin{equation}\label{mQ}
\tilde m^2=-2b_1G^2Q^2.
\end{equation}
This relation tells us that $b_1$ must always be less than zero and particularly for $b_1=-\frac{1}{2}$, one can obtain the relation $\tilde m^2=G^2Q^2$ similar to the one in the case of the Reissner-Nordstrom black hole in Einstein gravity, which is also obvious from (\ref{hLa2}).

The thermodynamics of the other case $a_1=0$ is similar to the case above in which $a_2=0$. In this case, the metric function $h(r)$ and the scalar aether field $\p(r)$ become
\begin{eqnarray}
&&h(r)=1-\frac{\La}{3}r^2-\frac{2a_2^2b_2} {r^{1-q}}-\frac{2\tilde{m}}{r},\label{hLa3}\\
&&\p(r)=\frac{a_2}{r^{(1-q)/2}}.\label{phiLa3}
\end{eqnarray}
This time, defining $a_2=GQr_0^{-(q+1)/2}$, where $Q$ is the NAT \ql charge" again, we can write (\ref{hLa3}) and (\ref{phiLa3}) as
\begin{eqnarray}
&&h(r)=1-\frac{\La}{3}r^2-\frac{2G^2Q^2b_2} {r^2}\l(\frac{r_0}{r}\r)^{-(q+1)}-\frac{2\tilde{m}}{r},\label{}\\
&&\p(r)=\frac{GQ}{r}\l(\frac{r_0}{r}\r)^{-(q+1)/2}.\label{}
\end{eqnarray}
At the event horizon location $r_0$, however, we obtain the same equations (\ref{hLa2}) and (\ref{phiLa2})
\begin{eqnarray}
&&h(r_0)=1-\frac{\La}{3}r_0^2-\frac{2G^2Q^2b_2} {r_0^2}-\frac{2\tilde{m}}{r_0}=0,\label{}\\
&&\p(r_0)=\frac{GQ}{r_0}.\label{}
\end{eqnarray}
The rest goes on like in the case of $a_2=0$; the only difference is that $b_1$ must be replaced by $b_2$ in all the equations (\ref{deltaS})-(\ref{mQ}).

\section{Null and Timelike Geodesics}

\subsection{Circular Orbits}

Here, we study the circular orbits at the equatorial plane, i.e $\theta=\frac{\pi}{2}$ , for the metric (\ref{hLa}) with $a_2=0$.
Accordingly, we have two Killing vector fields $K^\mu=\left( \partial_t \right)^\mu=(1,0,0,0)$ and $R^\mu=\left( \partial_\varphi \right)^\mu=(0,0,0,1)$ corresponding to
the conserved energy $E=-K_\mu \frac{dx^\mu}{d\sigma}$ and conserved angular momentum $L=R_\mu \frac{dx^\mu}{d\sigma}$, respectively, where $\sigma$ is an affine parameter
along the geodesics.
Then, regarding the metric, the energy and  angular momentum magnitude
of the orbiting body are given by
\begin{equation}\label{EL}
E=h\left( \frac{dt}{d\sigma}\right), ~~~~L=r^2\left( \frac{d\varphi}{d\sigma}\right).
\end{equation}
On the other hand, using the geodesics equation $g_{\mu\nu}\frac{dx^\mu}{d\sigma}\frac{dx^\nu}{d\sigma}=\epsilon$,
where $\epsilon=0$ and $-1$ denote the null and timelike geodesics, respectively,
we obtain
\begin{equation}
-h^2\left( \frac{dt}{d\sigma}\right)^2+\left( \frac{dr}{d\sigma}\right)^2+h\left[r^2\left( \frac{d\varphi}{d\sigma}\right)^2 -\epsilon\right]=0.
\end{equation}
Using the energy and angular momentum (\ref{EL}), we arrive at
\begin{equation}\label{em*}
\frac{1}{2}\left( \frac{dr}{d\sigma}\right)^2+ \mathcal{V}=\mathcal{E},
\end{equation}
where $\mathcal{E}=\frac{E^2}{2}$ and the potential $\mathcal{V}$ reads as
\begin{equation}
\mathcal{V}=\frac{1}{2}h\left(\frac{L^2}{r^2}-\epsilon  \right).
\end{equation}
Substituting the metric function $h$ in (\ref{hLa}), we find the potential as
\begin{equation}
\mathcal{V}=-\frac{\epsilon}{2}+\frac{\epsilon \tilde m}{r}+\frac{L^2}{2r^2}-\frac{\tilde
m
L^2}{r^3}-\frac{1}{6}\La L^2+\frac{1}{6}\epsilon\La r^2+\frac{\epsilon a_1^2 b_1 }{r^{1+q}}-  \frac{a_1^2 b_1 L^2}{r^{3+q}},
\end{equation}
where the first four terms are the standard terms as in GR \cite{Carroll}, and the last four terms are the new
correction terms by the cosmological constant and aether field, respectively.
In Figure 1, we have plotted the potential function $\mathcal{V}$ versus $r$ for some
sets of $q, \, L$ and $a_1^2 b_1$ parameters for the massive and massless
particles, respectively.
For each set of parameters,  one can see that in general the deviation of the potential
$\mathcal{V}$ from GR potential for the massive particles
is more than for the massless particles. For both the massive and massless
cases, by increasing $q$, the potential
tends to GR. However, by increasing $L$, the potential increases and deviates more
from GR. For $b_1>0$, the potential decreases by increasing $a_1^2 b_1$
values and vice versa.
 \\
\begin{figure}\label{pot}
\centering
\includegraphics[scale=0.59]{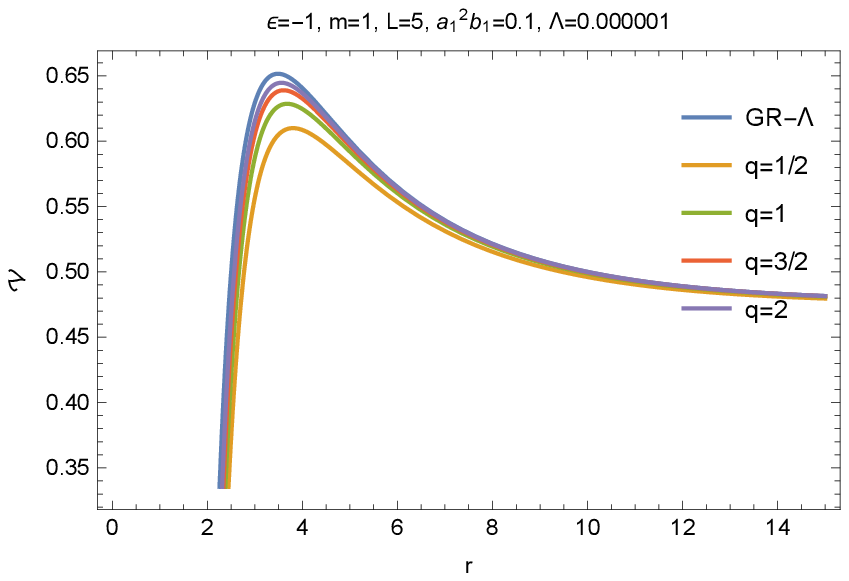}
\includegraphics[scale=0.59]{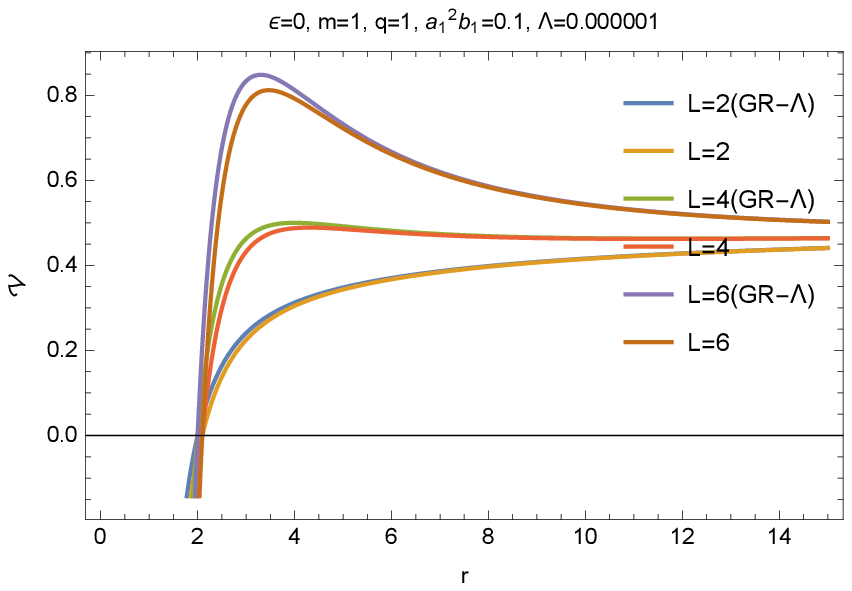}
\includegraphics[scale=0.59]{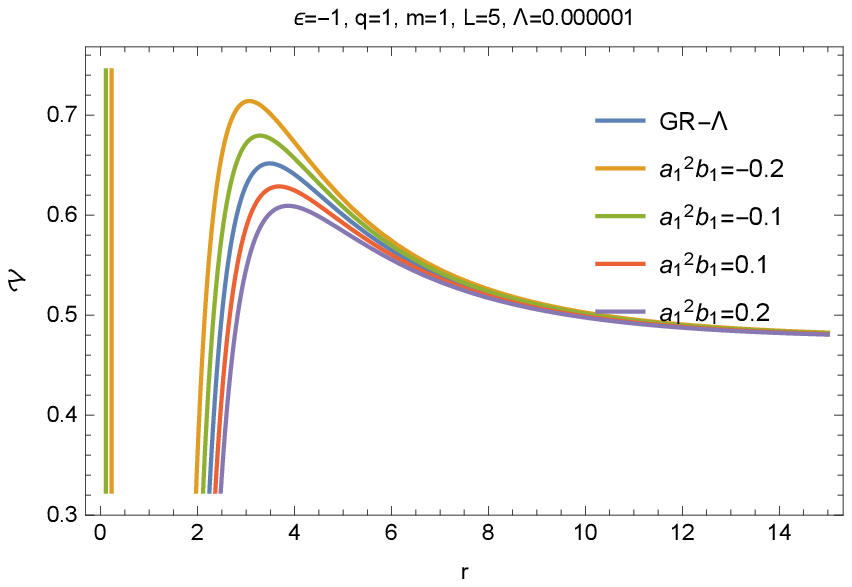}
\includegraphics[scale=0.59]{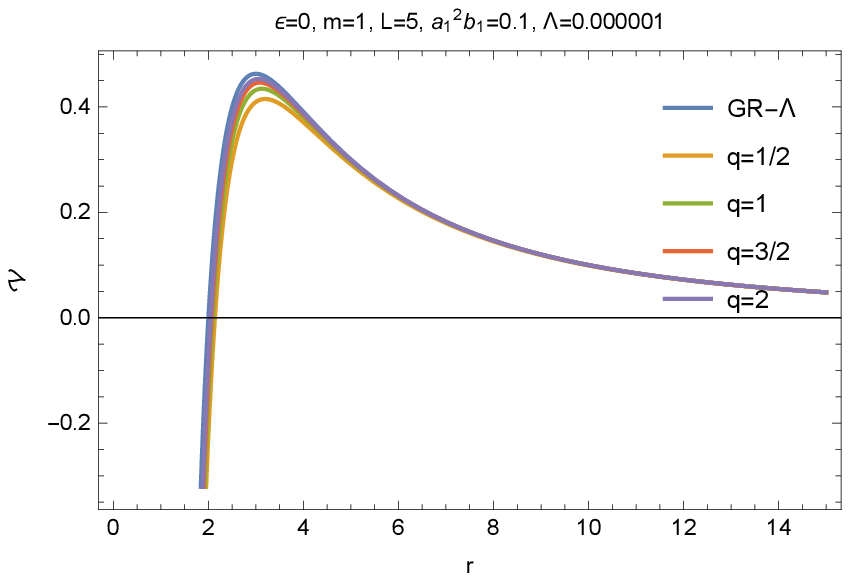}
\includegraphics[scale=0.59]{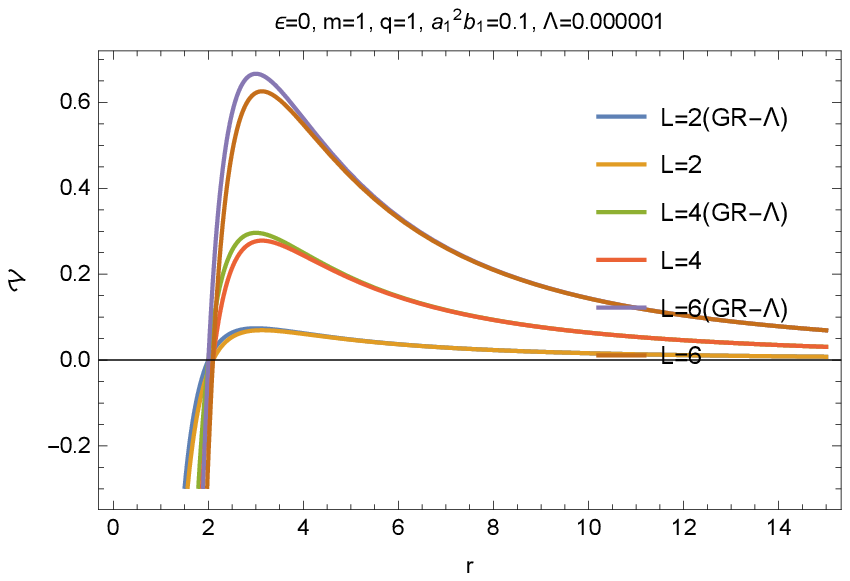}
\includegraphics[scale=0.59]{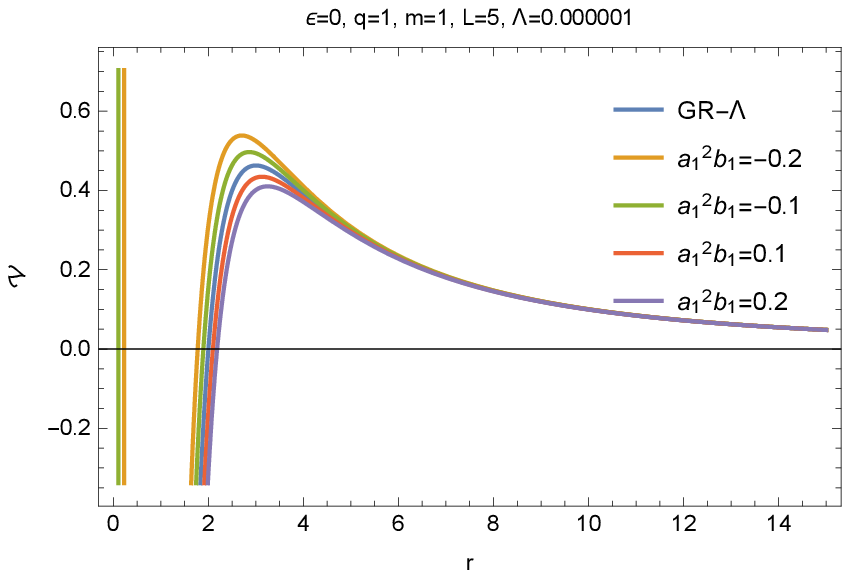}
\caption{The upper and  lower plots are denoting the potential $\mathcal{V}$ for some typical values of the parameters for the massive
and massless particles, respectively. }
\end{figure}
%
%
The circular orbits can be obtained as the
radii where the potential is flat, i.e  $\frac{d\mathcal{V}}{dr}\mid_{r=r_{c}}=0$. Here $r_c$ denotes the circular orbits.
Then, the equation governing the circular orbits can be obtained as
\begin{equation}\label{geo}
-\frac{\epsilon \tilde m}{r^{2}_c}-\frac{L^2}{r^{3}_{c}}+\frac{3\tilde m L^2}{r^{4}_c}+\frac{1}{3}\epsilon\La r-\frac{\epsilon(1+q) a_1^2 b_1 }{r^{2+q}_c}+ \frac{(3+q)a_1^2 b_1 L^2}{r^{4+q}_c}=0.
\end{equation}
For the GR limit by turning off the cosmological constant and aether field ($\La=0$ and $a_1=0$), we arrive at
\begin{equation}
-L^2 r_c +3\tilde m L^2 -\epsilon \tilde m r^{2}_c=0,
\end{equation}
which admits the following solutions for the massless and massive particles
respectively
\begin{equation}\label{grg}
  \left\{\begin{array}{ll}
         \displaystyle \mbox{ $\epsilon=0$:}&r_c=3\tilde m,\\
         &\\
         \displaystyle \mbox{ $\epsilon=-1$:}&r_{c^\pm}=\frac{L^2 \pm \sqrt{L^4 -12\tilde m ^2 L^2}}{2\tilde m}.
\end{array} \right.
\end{equation}
 In the presence of the cosmological constant and aether field, the equation (\ref{geo}) for the null
geodesics reduces to
\begin{equation}
r_c -3\tilde m -\frac{(3+q)a_1^2 b_1}{r^{q}_c}=0,
\end{equation}
where one can see that cosmological constant does not contribute for the
null
geodesics but the aether field does as the last
term. Here, one may consider the particular case $q=1$. This case has two
solutions as
\begin{equation}
r_{c^\pm}=\frac{3\tilde m}{2}\pm \frac{3\tilde m}{2}\sqrt{1 +\frac{16a_1^2 b_1}{9\tilde
m^2}}.
\end{equation}
Considering $9\tilde m^2\gg 16a_1^2 b_1$, we have
 \begin{equation}\label{}
 r_{c^\pm}\simeq \left\{\begin{array}{ll}
         3\tilde m +\frac{4}{3}\frac{a_1^2 b_1}{\tilde m},\\
         \\
         \displaystyle -\frac{4}{3}\frac{a_1^2 b_1}{\tilde m}.
\end{array} \right.
\end{equation}
Then, the second solution is a physical orbit only for $b_1<0$.
Thus, in contrast to GR which has only one null circular orbit as
in (\ref{grg}), in the presence of aether field  for $b_1<0$, there are two null circular orbits in which
the radius of the outer one is smaller than GR. For $b_1>0$, there is only one null circular orbit greater than the one in GR.

For the case of timelike circular orbits, solving Eq.\,(\ref{geo}) for a generic
$q$ is impossible. Thus, one may consider the particular case of $q=1$ where
the resulted equation will be a 6th order equation for $r_c$ as
(\ref{geo}) reduces to
\begin{equation}
L^2 r^{2}_{c} -3\tilde m L^2 r_{c} -\tilde m r^{3}_c +\frac{1}{3}\La r_c^6-4a_1^2 b_1 L^{2}-2a_1^2b_1  r^{2}_c=0.
\end{equation}
Finding the general real and positive solutions to this equation is not easy
task. However,
for realizing the effect of aether field, one may consider $\La=0$ and
$r_c>2\tilde m\gg 2a_1^2 b_1$ in the equation (\ref{geo}) which leads to
\begin{equation}
L^2 r_c -3\tilde m L^2 -\tilde m r^{2}_c -2a_1^2b_1  r_c=0.
\end{equation}
This equation has two
solutions as
\begin{equation}\label{rc}
r_{c^\pm}=\frac{L^2-2  a_1^2b_{1}}{2\tilde m}\pm \frac{\mid L^2-2  a_1^2b_{1}\mid}{2\tilde m}\sqrt{1 -\frac{12\tilde m^2}{L^2\left(1-\frac{2  a_1^2 b_{1}}{L^2}\right)^2}}.
\end{equation}
Following Carroll \cite{Carroll} for large $L$ values, we obtain
 \begin{equation}\label{}
 r_{c^\pm}\simeq \left\{\begin{array}{ll}
         \frac{L^2}{\tilde m}\left( 1-\frac{2  a_1^2 b_{1}}{L^2}\right),\\
         \\
         \displaystyle \frac{3\tilde m}{\left(1-\frac{2 a_1^2 b_{1}}{L^2}\right)} ,
\end{array} \right.
\end{equation}
where we have considered $1-2  a_1^2 b_{1} >0$.
Then, one can see that the aether field changes the inner and outer circular
orbits of massive particles in GR given by $r_{c^-}=3\tilde m$ and $r_{c^+}=\frac{L^2}{\tilde m}$, respectively. Accordingly, for $b_1<0$, the outer and inner circular
orbits will be larger and smaller, respectively, relative to GR and vice versa. For
\begin{equation}\label{l}
12 \tilde m^2=L^2\left(1-\frac{2 a_1^2 b_{1}}{L^2}\right)^2,
\end{equation}
these orbits coincide at
\begin{equation}\label{r6m}
r_c=\frac{6\tilde m}{1-\frac{2  a_1^2 b_{1}}{L^2}}.
\end{equation}
One can see from (\ref{r6m}) that the aether field changes the smallest possible
circular orbit for the massive particles as  $r_c=6\tilde
m$ in GR. 

\subsection{Perihelion Precession}

The perihelion precession represents that non-circular orbits are not
perfect closed ellipses.
To derive it, one should obtain the evolution of the radial coordinate $r$
as a function of angular coordinate $\varphi$, i.e. $r=r(\varphi)$. To do this,
 using (\ref{EL}) we write the equation (\ref{em*}) in the following form
\begin{equation}\label{em**}
\frac{1}{2}\left( \frac{dr}{d\varphi}\right)^2 \left(\frac{L}{r^2}  \right)^2+ \mathcal{V}=\mathcal{E}.
\end{equation}
For more convenience, we introduce a new variable as $x=\frac{1}{r}$. Then,
the above equation takes the following form
\begin{equation}\label{em***}
\frac{1}{2}\left( \frac{dx}{d\varphi}\right)^2 + \mathcal{\tilde V}(x)=\frac{\mathcal{E}}{L^2},
\end{equation}
where
\begin{equation}\label{poti}
\mathcal{\tilde V}(x)=\frac{1}{2}\left(1-2\tilde m x +\frac{\La}{3x^2}-2a_1^2 b_1x^{1+q} \right)\left(x^{2}-\frac{\epsilon}{L^2}  \right).
\end{equation}
Then, Eq.(\ref{em***}) for the timelike geodesics becomes
\begin{equation}
\frac{d^2 x}{d\varphi^2}+x=\frac{\tilde
m}{L^2}+3\tilde m x^2 +\frac{\La}{3L^2 }\frac{1}{x^3} +\frac{a_1^2 b_1(1+q)}{L^2}x^q+a_1^2b_1(q+3)x^{q+2}.
\end{equation}
This equation is the master equation for the perihelion precession in the
context of  NAT for generic $q$ and $b_1$ parameters.
Analytically solving this equation for generic $q$ is not an easy task and one may consider specific cases.
For the case of $q=1$, this equation reduces to

\begin{equation}\label{master}
\frac{d^2 x}{d\varphi^2}+x=\frac{\tilde
m}{L^2}+3\tilde m x^2 +\frac{\La}{3L^2 }\frac{1}{x^3}+\frac{2a_1^2 b_1}{L^2}\,x+4a_1^2b_1x^3.
\end{equation}
Then, in comparison to the Newtonian gravity possessing the equation $\frac{d^2 x}{d\varphi^2}+x=\frac{\tilde
m}{L^2}$, one can realize the  GR, cosmological constant and
aether field corrections, respectively. One can show that this equation admits
the following solution \cite{Hu}
\begin{eqnarray}
x(\varphi)&=&\frac{\tilde m}{L^2}\left[1+e \cos(\varphi) \right]+\frac{3\tilde m^3}{L^4}\left\{1+
e\varphi \sin(\varphi)+ \frac{e^2}{2} \left[1-\frac{1}{3}\cos(2\varphi)\right]\right\}\nonumber\\
&&+\frac{\La L^4}{3\tilde m^3}\left[1-\frac{3}{2}e \varphi\sin(\varphi) \right]+\frac{2
\tilde m a_1^2 b_1}{L^4}\left[1+\frac{1}{2}e\varphi \sin(\varphi) \right]\nonumber\\
&&+\frac{4\tilde m^3 a_1^2 b_1}{L^6}\left\{1+\frac{3}{2}e\varphi \sin(\varphi) + \frac{3e^2}{2} \left[1-\frac{1}{3}\cos(2\varphi)\right]\right\},
\end{eqnarray}
where  the first term is the solution for the Newtonian gravity with the eccentricity parameter $e$, and the other
terms are the corrections by GR, cosmological constant and aether field.
 Neglecting the higher order terms of the small eccentricity parameter $e$ and using the conditions $\frac{\tilde m^2}{L^2}\ll1$, $\frac{2a_1^2 b_1}{L^2}\ll1$,
one can rewrite the above equation as
\begin{eqnarray}
x(\varphi)&\simeq&\frac{\tilde m}{L^2}\left\{1+e \cos\l[\left(1-\zeta \right)\varphi\r] \right\},
\end{eqnarray}
where
\begin{equation}
\zeta=\frac{3 \tilde m^2}{L^2}-\frac{\La L^6}{2 \tilde m^4}+\frac{a_1^2 b_1}{L^2}.
\end{equation}
Then, during each orbit of the planet, there is a perihelion advance given
by \begin{equation}\label{peri}
\Delta \varphi=2\pi \zeta=2\pi \left(\frac{3 \tilde m ^2}{L^2}-\frac{\La L^{6}}{2\tilde m^4}+\frac{a_1^2 b_1}{L^2}\right).
\end{equation}
One can rewrite this relation by converting $L$ to the geometric quantities of
each orbit.  For this end, using the relation governing ordinary ellipses
as
\begin{equation}
r(\varphi)=\frac{(1-e^2)a}{1+e\cos(\varphi)},
\end{equation}
where $a$ is the semi-major axis, one can obtain the angular momentum as
 \begin{equation} \label{l2}
 L^2\approx \tilde m (1-e^2)a.
 \end{equation}
 Then, by substituting (\ref{l2}) in (\ref{peri}), we obtain \begin{equation}\label{peri1}
\Delta \varphi=\frac{6\pi \tilde m}{(1-e^2)a}\left[1-\frac{\La (1-e^2)^4
a^4}{6\tilde m^2}+\frac{a_1^2 b_1}{3\tilde m ^2}\right].
\end{equation}
Here, the correction term by the aether field  is exactly same as the one we
previously obtained  in (\ref{deltaphi1}) by using the post-Newtonian approximation.
It is seen  that for $\La>0$ and $b_1<0$, we always have less perihelion precession
relative to GR. However, for $\La>0$ and $b_1>0$, depending on the value of contributions by aether field and cosmological constant, we may
have more or less precession.
Also, there is an interesting case for $b_1>0$ in which the cosmological constant and
aether fields cancel out the effect of each other, i.e for $\La (1-e^2)^4
a^4=2a_1^2 b_1$,  leading to the same precession  as in GR.
\subsection{Light Deflection}
To obtain the deflection angle of null geodesics, we set $\epsilon=0$ in
the potential $\mathcal{V}$ in (\ref{poti}). Then, the equation governing
null geodesics takes the form of
\begin{equation}
\frac{d^2 x}{d\varphi^2}+x=3\tilde m x^2 +a_1^2b_1(q+3)x^{q+2},
\end{equation}
which shows that similar to the closed null geodesics in section 8.1, the
cosmological constant does not contribute to the light deflection angle.
However, the aether field contributes. Considering the case of $q=1$, this equation reduces to

\begin{equation}\label{master}
\frac{d^2 x}{d\varphi^2}+x=3\tilde m x^2 +4a_1^2b_1x^3,
\end{equation}
which has the following solution \cite{Hu}
\begin{equation}\label{dfg}
x(\varphi)=\frac{1}{r_0}\sin(\varphi)+\frac{\tilde m}{r_0^2}\left[1-\cos(\varphi)\right]^2 +\frac{a_1^2
b_1}{2r_0^3}\left[ -3\varphi \cos(\varphi) +\frac{1}{4}\sin(3\varphi) \right],
\end{equation}
where the first term represents a straight line in polar coordinates $(x,
\varphi),$ and
$r_0$ denotes the distance of closest approach of the light from
the gravitational center.
Then, the second and third terms denote the GR and aether field contributions
to the light deflection angle, respectively. The light deflection angle,
say $\xi$, can be
obtained using the condition $x(\pi +\xi)=0$ in  (\ref{dfg}) as
\begin{equation}
\xi\simeq\frac{4\tilde m}{r_0}+\frac{3\pi}{2}\frac{a_1^2 b_1}{r_0^2},
\end{equation}
where we have used the approximation relations $\sin(\pi +\xi)\simeq -\xi$ and $\cos(\pi +\xi)\simeq -1$ and dropped higher order terms in $\tilde m$
and $a_1^2 b_1$. Here, one realizes that depending on the sign of the aether field parameter $b_1$, the light deflection can be more or less than the GR value given by the above first term. For $b_1<0$, the aether field decreases the light deflection angle relative to the Schwarzschild case in GR. This is similar to the effect of charge  in the Reissner-Nordstr{\"o}m solution \cite{Hu, PRD}.

\section{Conclusion}

In this work, we investigated the properties of the black hole solutions found in NAT \cite{gsen1} which is a vector-tensor theory of gravity with the vector field being null and defining the aether field at each point of the spacetime. We first reviewed the Newtonian limit of the theory and showed that the Poisson equation is recovered at the linear order in the gravitation constant $G$ of the theory which, depending on the form of the null vector, is related to the Newton's constant $G_N$ by a scaling factor. We also reviewed the exact spherically symmetric static solutions in NAT and extracted the post-Newtonian parameters $\b,\g$ from these solutions when $\La=0$. In GR, these parameters are $\b=\g=1$ and in NAT, for $q=0$, we have the same values because the solution is the usual Schwarzschild metric in this case. However, for solutions with $q>0$, taking $a_2=0$ for simplicity, we found that $\b=1-\frac{a_1^2b_1}{\tilde{m}^2}$ and $\g=1$, meaning that, at the post-Newtonian order, the aether does not contribute to the light deflection expression, which is determined only by $\g$, while it contributes to the perihelion advance expression, which is determined by both $\b$ and $\g$. Since the perihelion advance differs from the GR value by the term ($\frac{a_1^2b_1}{3\tilde{m}^2}$) where $b_1\e=\frac{1}{8}[c_3-3c_2+c_{23}q]$ [see Eq. (\ref{deltaphi1})], the effect of the null aether is such that the GR value for the perihelion advance of planets is increased (for $b_1>0$) or decreased (for $b_1<0$). That is to say, solar system observations can be used to put some constraints on the parameters of the theory.

We also studied the exact static black hole solutions in NAT. We observed that, depending on the parameter $q$, there is a large class of black hole solutions in the theory and showed, by calculating the curvature scalars Ricci and Kretschmann, that all the these solutions are singular only at $r=0$. These black holes possess in general multiple event horizons and the locations of these horizons are dependent on the parameters $(q, \La, a_1, a_2, b_1, b_2, \tilde{m}, m)$ and the relations between them. There are also extreme cases in which some or all of the event horizons coincide. To determine the mass parameters of these solutions, we calculated the ADM mass of the asymptotically flat black holes and showed that, just like the mass parameter $m$ in the case $q=0$, the mass parameter in the case $q>0$ reads $\tilde{m}=GM_{ADM}$, where $G$ is the gravitational constant appearing in the theory.

In the thermodynamics discussion of the NAT black holes, we carried out a generic analysis in which the cosmological constant is nonzero. First, defining the NAT \ql charge" appropriately, we showed that the horizon condition $h(r_0)=0$ and the scalar aether field $\p(r_0)$ at the horizon become similar to the ones of the Reissner-Nordstr\"{o}m-(A)dS black hole in GR, independently of the values of the parameter $q$. Then we obtained the first law of thermodynamics in which the contribution of the aether field appears as $V_\p\d Q$, where $V_\p=-2b_1\p(r_0)$ with $\p(r_0)=\frac{GQ}{r_0}$ and $Q$ is the NAT charge. Therefore, for consistency, it turns out that $b_1=-1/2$ to recover the standard form of the first law.

Lastly, we studied both the null and timelike geodesics in the NAT black hole geometries. We explicitly derived the general expression for the effective potential governing the motion of the particles in the gravitational field including the correction terms due to the cosmological constant and the aether field. As is shown in Fig. (1), depending on the values of $(q, L, a_1^2b_1)$, it turns out that the deviation of the potential from the GR value is more in the case of massive particles than in the case of massless particles. In addition, by increasing $q$, the potential tends to the GR one, while, by increasing $L$, it deviates more from the GR value for both the massive and massless particles. We also obtained the general equation governing the location of the circular geodesics for both massive and massless particles to which there is no contribution from the cosmological constant for the null geodesics as in GR. For specifically $q=1$, we showed that, in contrast to GR possessing only one null circular orbit, there are two circular orbits in the presence of the aether field for $b_1<0$, and of them, the outer one has a smaller radius than that of the one in GR. For $b_1>0$, on the other hand, there is always only one circular orbit the radius of which is greater than the one in GR. In the case of timelike geodesics, again for $b_1<0$, there are two different circular orbits: the outer and the inner ones are, respectively, larger and smaller that the ones in GR. As a particular case, when these circular orbits coincide, the aether field makes the location smaller than in GR if $b_1<0$. We further studied the perihelion advance of massive particles in this context. We explicitly calculated the contributions of the cosmological constant and the null aether field and showed that, when $\La=0$, the aether field contribution is the exactly the same as the one obtained in the post-Newtonian order. Finally, we investigated the issue of light deflection angle. We showed that the cosmological constant does not contribute to the light deflection angle. However, the aether field contributes in which, depending on the sign of the $b_1$ parameter, the light deflection can be more or less than in GR. Indeed, for $b_1<0$, the aether field decreases the light deflection angle relative to the Schwarzschild solution in GR.

NAT is a new modified theory of gravity recently introduced \cite{gsen1}. So far, we have investigated this theory from various respects: Newtonian limit, spherically symmetric solutions, black holes, thermodynamics, circular geodesics, flat cosmological solutions, exact plane waves, etc. But there are some open problems regarding, for example, the stability of the theory, linearized waves, rotating black holes, generic cosmological solutions, inflationary cosmologies and etc. Therefore, to gain more understanding on the internal structure and dynamics of NAT, one needs to further investigate and pose analytical solutions to the theory.


\section*{Acknowledgements}

This work is partially supported by the Scientific and Technological Research Council of Turkey (TUBITAK).

\end{document}